\documentclass[prx,twocolumn,showpacs,superscriptaddress,preprintnumbers,amssymb]{revtex4-2}
\usepackage{graphicx}
\usepackage{float}
\usepackage{color}
\usepackage{xcolor}
\usepackage{dcolumn}
\usepackage{bm}
\usepackage{physics}
\usepackage{comment}
\usepackage{svg}
\usepackage{placeins}
\usepackage{hyperref}
\usepackage{soul}
\usepackage{feynmp-auto}
\usepackage{tikz}
\usepackage[normalem]{ulem}
\usetikzlibrary{decorations.pathmorphing}
\usetikzlibrary{arrows.meta,decorations.markings}

\newcommand{\beq}{\begin{equation}}
\newcommand{\eeq}{\end{equation}}
\newcommand{\beqn}{\begin{eqnarray}}
\newcommand{\eeqn}{\end{eqnarray}}

\newcommand{\Eq}[1]{Eq.~(\ref{#1})}

\newcommand{\cD}{ {\cal D} }

\newcommand{\SU}{\mathrm{SU}}

\newcommand{\Sp}{\mathrm{Sp}}

\newcommand{\figref}[1]{Fig.~\ref{#1}}
\newcommand{\sfigref}[2]{Fig.~\hyperref[#1]{\ref{#1}#2}}

\newcommand{\Ecal}{\mathcal E}

\providecommand{\tr}{\operatorname{tr}}

\providecommand{\ket}[1]{|#1\rangle}
\providecommand{\bra}[1]{\langle #1|}

\newcommand{\one}{\mathbf 1}

\definecolor{rededge}{rgb}{0.776,0.243,0.208}
\definecolor{blueedge}{rgb}{0.153,0.463,0.722}
\graphicspath{{figures/}}
\hypersetup{colorlinks=true,linkcolor=blue,citecolor=blue,urlcolor=blue}

\begin{document}

\title{Threshold-Free Symmetry-Enriched Topological Quantum Memory}
\date{\today}

\author{Abhijat Sarma}

\affiliation{Department of Physics, University of California,
Santa Barbara, CA 93106, USA}

\author{Ruochen Ma}

\affiliation{Kavli Institute for Theoretical Physics, University of California, Santa Barbara, CA 93106, USA}

\author{Cenke Xu}

\affiliation{Department of Physics, University of California,
Santa Barbara, CA 93106, USA}

\begin{abstract}

We show that non-Abelian symmetry enrichment can protect a toric code memory against incoherent electric noise at any bond error probability $p\in[0,1]$. For a toric code on the honeycomb lattice coupled to $\Sp(N)$ rotors, with electric anyons carrying the fundamental representation, we construct an efficient minimum-weight decoder whose logical error probability decays exponentially with system size for $N\ge40$, a conservative sufficient bound. The same decoder also rules out strong-to-weak spontaneous symmetry breaking in the fundamental channel of $\Sp(N)$ under the corresponding local decoherence in two dimensions. The construction extends to other symmetries, including $\SU(N)$.

\end{abstract}

\maketitle

\section{Introduction}

A topological quantum memory (TQM) encodes coherent quantum information in its topological sectors. However, a TQM is not generically invincible to noise and decoherence. Under decoherence, there is usually a finite threshold of decoherence strength beyond which the stored information is lost and can no longer be recovered. Such decodability phase transitions are well-established in the toric code~\cite{dennis2002,Lee_2023,Fan2023,bao2026}, and other examples of TQMs~\cite{katzgraber2009,andrist2011,ducloscianci2013,hutter2015,wootton2014,varona2020,heralding1,heralding2,chen2024,sajith2026,weinstein2026,fusiontree2026,sagar2026}. Heuristically, a TQM becomes undecodable when noise {\it incoherently proliferates} anyons, which maps to a {\it coherent condensation} of anyon bound states between the ket and bra sectors of the density matrix~\cite{Lee_2023,Fan2023,bao2026}. It was hence realized that the decodability transition of TQM is also closely related to a novel structure of spontaneous symmetry breaking, now known as the strong-weak spontaneous symmetry breaking (SW-SSB)~\cite{Lee_2023,Fan2023,bao2026,lessa2025strong,sala2024spontaneous,gu2025spontaneous,huang2025hydrodynamics,zhang2025strong,zhang2024fluctuation,liu2025diagnosing,guo2025strong,weinstein2025efficient,hauser2026,localSWSSB1,localSWSSB2,localSWSSB3,sarma2026, wang2026}. This correspondence allows us to study the decodability transition, an inherently information-theoretic transition, using the familiar Landau-Ginzburg framework of condensed matter physics. In particular, the decodability transition of the toric code is mapped to the $2d$ classical random bond Ising model~\cite{dennis2002,Lee_2023,Fan2023,bao2026}. In fact, incoherent proliferation of particles and anyons driven by noise is often mapped to a finite temperature $2d$ classical statistical mechanics model. These observations provided us with very helpful insights on the following question:

\medskip

{\it Is there a TQM robust against decoherence for all decoherence strength $p \in [0,1]$? } 

\medskip

This question is equivalent to 

\medskip

{\it Is there a TQM whose anyons cannot be incoherently proliferated by noise? } 

\medskip

The discussions in the previous paragraphs point to one promising answer: if the anyons carry a nonabelian continuous symmetry, the Mermin-Wagner theorem states that this symmetry cannot be spontaneously broken at any finite temperature in $2d$ classical statistical mechanics models. Though a tractable statistical mechanics mapping is not always available, this suggests that nonabelian continuous symmetries can potentially prohibit anyon proliferation, and the system should remain decodable for any $p \in [0, 1]$. 

Symmetry enriched TQMs have been discussed in recent literature~\cite{temkin2025,fusiontree2026,ryan2026}, and Ref.~\cite{temkin2025} was directly motivated by the Mermin-Wagner theorem. In these examples symmetry indeed improves the decoding threshold $p_c$ of the topological order, but a fully decodable TQM for the entire range $p \in [0,1]$ has not yet been identified, and an exact relation between the decodability of $2d$ TQM and the MW theorem remains elusive. In this work we demonstrate that large enough symmetries indeed ensure decodability: for $\SU(N)$ and $\Sp(N)$ enriched TQM, for $N$ greater than a finite critical value $N_c$, the system remains decodable by a polynomial-time decoder and strongly symmetric for any $p \in [0,1]$. Our construction only relies upon the fact that the symmetry group at large-$N$ heralds the degree of the error chain perfectly at each vertex, which is true for many groups with sufficiently rich representation theory.

\section{Symmetry enriched Toric code}

\label{sec:model}

We consider a toric-code TQM with $\Sp(N)$ rotor matter, where an elementary error creates two electric anyons carrying an $\Sp(N)$ fundamental representation at each endpoint. Throughout, $N$ is even and denotes the dimension of the fundamental representation, so that $\Sp(2) = \SU(2)$. We will assume a trivalent honeycomb lattice with periodic boundary conditions throughout.

The Hilbert space consists of qubits on each edge and an $\Sp(N)$ rotor on each vertex. The initial state is
\beq
\begin{gathered}
\ket{\mathrm{TC}}\otimes \prod_i\ket+_i,\\
\ket+_i=\int dg_i\,\ket{g_i},
\qquad \int dg_i=1,
\end{gathered}
\label{eq:initial}
\eeq
where $\ket{\mathrm{TC}}$ is an arbitrary toric-code ground state and $\ket+_i$ is an on-site $\Sp(N)$ singlet. 

For an edge $e=\{i,j\}$, define the $N^2$ Kraus operators, with $\alpha,\beta=1,\dots,N$,
\beq
\begin{aligned}
K^{\alpha\beta}_{ij} ={}& Z_{ij}\otimes\frac1{\sqrt N} \int dg_i\,dg_j\\
&\; D^f_{\alpha\beta}(g_i^{-1}g_j) \ket{g_i,g_j}\bra{g_i,g_j},
\end{aligned}
\label{eq:Kraus}
\eeq
where $D^f$ is the fundamental representation matrix of $\Sp(N)$. The channel on this edge is
\beqn
\Ecal_e(\rho)=(1-p)\rho+
p\sum_{\alpha,\beta}K^{\alpha\beta}_{ij}\rho K^{\alpha\beta\dagger}_{ij}.
\label{eq:channel}
\eeqn
This is a large-$N$ generalization of the channel considered in Ref.~\cite{fusiontree2026}. Since $D^f(u)$ is unitary,
\beqn
\sum_{\alpha,\beta}K^{\alpha\beta\dagger}_{ij}K^{\alpha\beta}_{ij}
=\frac1N\tr[D^{f\dagger}D^f]\,\one=\one.
\label{eq:normalization}
\eeqn

The full noise channel is the composition of these edge channels. Each edge is independently activated with probability $p$. Every Kraus operator also preserves the global strong $\Sp(N)$ symmetry, since $g_i^{-1}g_j$ is unchanged by $g_i\mapsto h g_i$ on every site. 

Here we discuss $\Sp(N)$ rotors rather than $\SU(N)$ because the channel for $\Sp(N)$ is \emph{undirected}:
\beqn
\sum_{\alpha,\beta}K^{\alpha\beta}_{ij}\rho K^{\alpha\beta\dagger}_{ij} = \sum_{\alpha,\beta}K^{\alpha\beta}_{ji}\rho K^{\alpha\beta\dagger}_{ji}.
\label{eq:undirected}
\eeqn
The fundamental and antifundamental representations of $\Sp(N)$ are equivalent, and exchanging the two endpoints only mixes the Kraus labels by a fixed unitary transformation. For $\SU(N)$ with $N > 2$, the fundamental and antifundamental are inequivalent, and the corresponding channel is directed. The direction information is not important to the discussion below and therefore our results also apply to $\SU(N)$, though we focus on $\Sp(N)$ for simplicity. 

\section{Infinite $N$: Perfect degree heralding}

\label{sec:largeN}

Let $E$ be the set of actual activated edges, and let $k_i$ be the number of these edges touching site $i$ (called the degree of $E$ at $i$). Each activated edge contributes one fundamental representation to each endpoint, so the representation measured at $i$ comes from $F^{\otimes k_i}$, the fusion of $k_i$ fundamental reps of $\Sp(N)$. 

At $N = \infty$, the measured onsite  $\Sp(N)$ representation (the syndrome) determines $k_i$ completely:

\begin{itemize}

\item $k_i = 0$: the on-site representation is a singlet.

\item $k_i=1$: the on-site representation is the fundamental $F$.

\item $k_i=2$: the representation is nontrivial and even under the $Z_2$ center of $\Sp(N)$. The two fundamentals can form a singlet, a symmetric representation, or a traceless antisymmetric representation. The singlet probability is exactly $1/N^2$, and therefore vanishes at large $N$. 

\item $k_i=3$: the representation is odd under the $Z_2$ center, but not a fundamental. At finite $N$, the probability for three fundamentals to fuse into another fundamental is exactly $3/N^2$, which again vanishes at large $N$.
\end{itemize}

The local syndromes determine much of the error configuration, therefore they serve the purpose as heralding discussed before~\cite{heralding1,heralding2}. In the large-$N$ limit, the syndromes completely determine the number of error edges passing through a site, therefore we have {\it perfect degree-heralding} in this limit. But even in this limit, the syndromes do {\it not} determine every error bond. 

We define a ``minimum-weight" decoder, which returns a candidate error string $E'$ of minimum length $|E'|$ fully consistent with the degree count $k_i$ at every vertex. This decoder always exists, and outputs $E'$ in polynomial time. In the perfect degree-heralding limit the minimum-weight condition is superfluous because $2|E'| = 2|E| = \sum_i k_i$, but it will be important in the next section. After applying the correction on $E'$, the remaining qubit operation is supported on 
\beqn
&& C = E \oplus E',
\cr\cr 
&& E \oplus E' \equiv (E\setminus E')\cup(E'\setminus E),
\label{eq:disagreement}
\eeqn
the edges belonging to exactly one of the two sets, and edges present in both sets cancel.

\begin{figure}[htbp]
\centering
\includegraphics[width=\columnwidth]{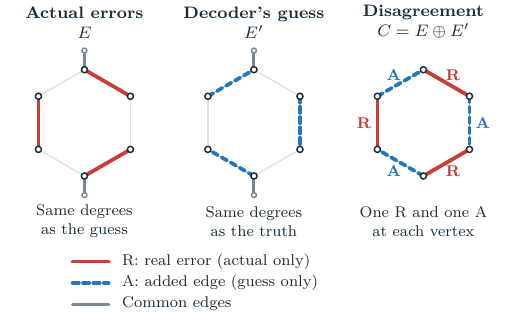}
\caption{Compare the actual errors $E$ with the decoder's guess $E'$. Common edges cancel, leaving the disagreement-loop $C=E \oplus E'$. The length of a disagreement loop counts both red and blue edges. 
}
\label{fig:disagreement}
\end{figure}

To describe this comparison, we use the colors in \figref{fig:disagreement}:
\begin{itemize}
\item \textcolor{rededge}{\textbf{$R$: real error, red.}} These edges belong to $E$ but not to $E'$: actual errors omitted by the decoder.
\item \textcolor{blueedge}{\textbf{$A$: Assumed but incorrect error, blue.}} These edges belong to $E'$ but not to $E$: edges assumed by the decoder where no error actually occurred.
\end{itemize}

The disagreement loop $C = E \oplus E'$ must form loops alternating between $R$ edges and $A$ edges because the true and guessed errors have the same degree at every vertex, as visualized in Fig.~\ref{fig:disagreement}. 
For a fixed loop of even length $\ell$, there are exactly two alternating activation patterns. Each contains $\ell/2$ $R$ edges and $\ell/2$ $A$ edges. Since the error edges are activated independently,
\beqn
\Pr(\text{a fixed loop is alternating})
=2[p(1-p)]^{\ell/2}.
\label{eq:alternating-probability}
\eeqn
Edges outside this loop are unrestricted. This probability is largest at $p=1/2$, where it equals $2^{1-\ell}$.

It remains to count the possible loop shapes. A simple winding loop on the torus lifts to a self-avoiding walk on the infinite honeycomb lattice. The number of rooted self-avoiding walks of length $\ell$ grows as $\mu^\ell$ up to subexponential factors, where~\cite{honeycomb}
\beqn
\mu=\sqrt{2+\sqrt2}\simeq1.847759<2
\label{eq:connective}
\eeqn
on the honeycomb lattice. The rest of the argument is therefore Peierls-like: there is a competition between a number of shapes growing with rate $\mu$ and a probability decreasing with rate $[p(1-p)]^{1/2}$, whose peak value is $1/2$ at $p = 1/2$. Let $L$ be the length of the shortest noncontractible cycle. A logical failure requires at least one noncontractible disagreement loop, hence a loop with $\ell\ge L$. Summing over shapes, starting sites, and all such lengths gives
\beqn
P_{\mathrm{fail}} \le e^{o(L)}\left(\mu \sqrt{p(1-p)} \right)^L
\longrightarrow0.
\label{eq:largeN-failure}
\eeqn
Here $e^{o(L)}$ denotes factors that grow more slowly than exponentially in $L$. Therefore in the thermodynamics limit, the failure rate of decoding vanishes. 

\section{Finite $N$: Imperfect degree heralding}

\label{sec:finiteN}


At finite $N$, the representation measurement still determines the parity of $k_i$ exactly. What can be lost is a pair of activated edges that combine into a singlet.

On the honeycomb lattice, the possible interpretations are:
\begin{itemize}
\item A measured singlet can mean $k_i=0$ or $k_i=2$.
\item A measured fundamental can mean $k_i=1$ or $k_i=3$.
\item Any other even representation fixes $k_i=2$, and any other odd representation fixes $k_i=3$.
\end{itemize}
We call the reductions $2\to0$ and $3\to1$ \emph{hidden pairs}.

\begin{figure}[htbp]
\centering
\includegraphics[width=\columnwidth]{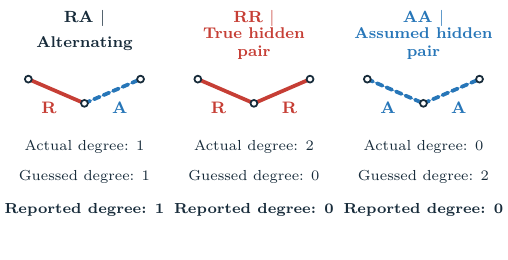}
\caption{The three local patterns on a disagreement loop. Both $RR$ and $AA$ interrupt alternation, but only $RR$ forces a hidden pair in the actual noise realization. 
}
\label{fig:interruptions}
\end{figure}

We again consider the ``minimum-weight" decoder which returns the smallest edge set $E'$ compatible with the measured syndromes. Minimizing $|E'|$ is equivalent to minimizing the number of assumed hidden pairs.
The parity of degree is still exact, so $C = E \oplus E'$ still consists of disjoint closed loops. However, these loops need not alternate perfectly. Due to hidden pairs, the alternation can be interrupted. There are three possible local patterns of interruptions, shown in \figref{fig:interruptions}:
\begin{itemize}
\item \textbf{$RA$: ordinary alternation.} The true configuration has one loop edge and the guess has the other. Their syndromes agree.

\item \textbf{$RR$: an actual hidden pair.} The true configuration has two more incident edges than the guess. The measurement must have hidden a pair in the actual state. If the common third edge is inactive, this is $2\to 0$; if it is active, this is $3 \to 1$.

\item \textbf{$AA$: a hidden pair assumed by the decoder.} The guess has two more incident edges than the truth. 

\end{itemize}

Now consider one disagreement loop. 
Let $n_{RR}$ and $n_{AA}$ be the numbers of $RR$ and $AA$ interruptions. Now count edge ends along the loop. Compared with an $RA$ vertex, every $RR$ vertex contributes one extra red link, every $AA$ vertex contributes one extra blue link. Thanks to the minimum-weight decoder we chose,
\beqn
n_{RR} - n_{AA} = n_R - n_A \ge 0
\label{eq:interruption-balance}
\eeqn
which follows from the fact that $E'$ has the minimal number of assumed hidden pairs. Thus at least half of all interruptions are $RR$ vertices and require actual hidden pairs.

The remaining input is a bound on the probability of several hiding events. Let $ 0 \le q \le 1 $ be a number such that, for every fixed activated-edge configuration $E$ and every specified set of $m$ vertices,
\beqn
\Pr(\text{all these $m$ vertices hide} \mid E)\le q^m.
\label{eq:joint-hiding}
\eeqn
We give an explicit evaluation of $q(N)$ in Section~\ref{sec:qbound}.

If a disagreement loop has $t$ interruptions, at least $t/2$ are actual hidden pairs. Their joint probability is at most $q^{t/2}=(\sqrt q)^t$. 
We first consider $p = 1/2$. For a fixed loop of length $\ell$, every specified red/blue assignment has activation probability $2^{-\ell}$. Choose $t$ of its vertices as interruptions and choose the first edge's color. At an ordinary vertex, switch color; at an interruption, keep the same color. This way all remaining colors are then fixed.

There are at most $2\binom{\ell}{t}$ assignments of the locations of the $t$ interruptions on a loop with length $\ell$. Some fail to close consistently or violate the minimum-edge condition, and including them only enlarges the bound. Summing over $t$ gives
\beqn
&& \Pr(\text{a fixed loop is a disagreement loop})\cr\cr
&& \le 2^{1-\ell}\sum_{t=0}^{\ell}\binom{\ell}{t}(\sqrt q)^t  = 2\left(\frac{1+\sqrt q}{2}\right)^\ell.
\label{eq:fixed-loop}
\eeqn

We can also prove the same bound for all $p$. Fix one red/blue assignment before taking probabilities. Its red edges must be activated, its blue edges must be inactive, and its $RR$ vertices must hide. Equation~\eqref{eq:joint-hiding} holds for every configuration outside the loop, so averaging over those outside edges gives the upper bound
\beqn
p^{n_R}(1-p)^{n_A}q^{n_{RR}}.
\label{eq:assignment}
\eeqn
The total number of interruptions is $t = n_{RR} + n_{AA}$. Together with \Eq{eq:interruption-balance}, this gives $2 n_{RR} = t + n_R - n_A$. Therefore
\beqn
p^{n_R}(1-p)^{n_A}q^{n_{RR}} &=& q^{t/2}[p(1-p)]^{n_A}[p\sqrt q]^{n_R - n_A} \cr\cr &\le & q^{t/2}(1/4)^{n_A}(1/2)^{n_R - n_A} \cr\cr &=&  2^{-\ell}(\sqrt q)^t,
\label{eq:all-p}
\eeqn 
provided $q\le1/4$. We used only $p(1-p)\le 1/4$, $p\sqrt q\le1/2$, $n_R \ge n_A$, and $\ell=n_R + n_A$. Thus \Eq{eq:fixed-loop} holds uniformly for all $p\in[0,1]$.

As in argument of the large-$N$ limit, count the loop shapes, sum over starting sites, and sum over lengths $\ell\ge L$. This gives
\beqn
P_{\mathrm{fail}} \le e^{o(L)} \left[\frac{\mu(1+\sqrt q)}2\right]^L.
\label{eq:finiteN-failure}
\eeqn
The failure probability therefore vanishes exponentially whenever
\beqn
\mu(1+\sqrt q)<2.
\label{eq:criterion}
\eeqn

The remaining task is to estimate and bound $q(N)$ for finite $N$. The appendix shows that \Eq{eq:joint-hiding} holds with
\beqn
q(N) = \frac{3N^{1/3}}{N(N-2)},
\qquad N \ge 4 .
\label{eq:qN}
\eeqn
The decoding criterion \Eq{eq:criterion} becomes
\beqn
q(N)<\left(\frac2{\sqrt{2+\sqrt2}}-1\right)^2.
\label{eq:numerical-criterion}
\eeqn
Every even $N \ge 40$ satisfies this criterion. Therefore
\beqn
q_c \sim 0.0068, \quad N_c \leq 40
\label{eq:Nc}
\eeqn
is a sufficient bound for the stated $\Sp(N)$ rotor channel, uniformly for all $p\in[0,1]$. The estimate is conservative and does not determine the sharp critical $N_c$, which is likely much less than $40$. The same arguments for $\SU(N)$ allow us to prove that $N_c \leq 31$ in that case. Further, these bounds apply to the particular minimum-weight decoder we chose. The optimal decoder likely also has $N_c$ much lower than the $N_c$ of this minimal-weight decoder. The decoder also tolerates imperfect irrep readout, assuming the readout of the degree-parity is exact. We will discuss imperfect readout in another work.

\section{SW-SSB}

\label{sec:rotor-decoding}

We now consider a pure rotor model without the $Z_2$ gauge field, obtained by omitting $Z_{ij}$ from \Eq{eq:Kraus}. Let $\widetilde K_{ij}^{\alpha\beta}$ denote the remaining rotor Kraus operators, and let $\rho$ be the state obtained by applying this noise channel to $\prod_i\ket+_i$. We define a new charge-decoding task: distinguishing $\rho$ from the state with an additional pair of fundamental charges inserted at $x,y$,
\beqn
\rho_{xy} = \sum_{\alpha,\beta} \widetilde K_{xy}^{\alpha\beta}\rho \widetilde K_{xy}^{\alpha\beta\dagger}.
\label{eq:rotor-pair-insertion}
\eeqn
This insertion reverses the center parity at $x,y$ and leaves the reduced state elsewhere unchanged.
A decoding channel $\cD$ outputs $0$ or $1$ to identify $\rho$ or $\rho_{xy}$, respectively. Define $a = \Pr(\cD(\rho)=1)$, $b=\Pr(\cD(\rho_{xy})=0)$, and $P_e=\max\{a,b\}$. 

We consider the fidelity $F(\rho,\sigma)=\tr\sqrt{\sqrt{\rho}\sigma\sqrt{\rho}}$. The monotonicity under the decoding channel gives
\beqn
F(\rho,\rho_{xy}) &\le& F(\cD(\rho),\cD(\rho_{xy})) \nonumber \cr\cr
&=&
\sqrt{(1-a)b}+\sqrt{a(1-b)}
\nonumber\cr\cr
&\le& 
2\sqrt{P_e}.
\label{eq:charge-decoding-fidelity}
\eeqn 
$F(\rho,\rho_{xy})$ is precisely the fidelity correlator that diagnoses the SW-SSB~\cite{lessa2025strong}. Thus an exponentially small charge-decoding error would imply an exponentially decaying fidelity correlator.

Our minimum-weight decoder can be straightforwardly generalized to this task. To construct $\cD$, measure the on-site $\Sp(N)$ irreps, retaining only the center parity at $x$ and $y$. For each hypothesis $h\in\{0,1\}$, where $h=0$ means no probe pair and $h=1$ means an additional fundamental pair at $x,y$, define
\begin{equation}
  m_h=\min_{E'\text{ compatible with }h}|E'|.
  \label{eq:charge-min-weight}
\end{equation}
A compatible $E'$ has enough incident legs to account for the measured irreps away from $x,y$, and matches the measured center parities at all sites after including the hypothesized probe legs at $x$ and $y$. In particular, under $h=1$, the required noise parity is reversed at both endpoints. The decoder then chooses the hypothesis with smaller $m_h$.

As before, a wrong answer requires $E\oplus E'$ to contain a self-avoiding path $P$ connecting $x,y$. Minimum-weight implies $n_R-n_A\ge0$, which gives $n_{RR}\ge n_{AA}-1$ along this open path. Repeating the preceding counting argument, with $r=d(x,y)$,
we obtain
\begin{equation}
  P_e(r)\lesssim e^{o(r)}
  \left[\frac{\mu(1+\sqrt{q(N)})}{2}\right]^r.
  \label{eq:charge-decoding-error}
\end{equation}
Thus the same condition $q(N)<q_c$, satisfied by all even $N\ge40$, guarantees exponentially decaying fidelity correlator for every $p\in[0,1]$, excluding SW-SSB detected by fundamental probes. For center-neutral probe irreps, fusion can hide the additional charges from the local irrep record, as discussed in Ref.~\cite{fusiontree2026}; this decoder argument does not settle their SW-SSB.

\section{Discussion}

In this work we demonstrate that toric code topological quantum memory enriched with $\Sp(N)$ and $\SU(N)$ symmetry under decoherence remains decodable for all $p \in [0,1]$, for $N > N_c$, inspired by the recent understanding of strong-weak spontaneous symmetry breaking. The critical $N_c$ estimated in this work is very conservative, which should be above the actual critical bound. We will determine the actual bound in future numerical work.

Effective models with large-$N$ symmetries may be realized in various platforms across condensed matter and cold atoms~\cite{PhysRevLett.81.5406,LiMaShiZhang1998SU4,VanDenBossche2001Plaquette,WuHuZhang2003SO5,Wu2006HiddenSymmetry,Gorshkov2010TwoOrbitalSUN,YamadaOshikawaJackeli2018SU4}. In particular, topological orders can be naturally realized in spin liquids, especially spin systems with large $\Sp(N)$ and $\SU(N)$ symmetries~\cite{wenZ2,subirZ2}. In the future we will also discuss stability of spin liquid states against decoherence and noise, whose gauge field is emergent, rather than built-in explicitly. Although the model considered in this work consists of a toric code coupled to rotors, our argument of decodability based on degree-heralding also applies to other formalisms with $\Sp(N)$ or $\SU(N)$ symmetry, including Schwinger-boson and fermionic-parton constructions commonly used in study of quantum spin liquids. 

\medskip

{\it --- Acknowledgment}: The authors thank Ehud Altman, Romain Vasseur, Ashvin Vishwanath, Yi-Zhuang You for very helpful discussions, and acknowledge the substantial contributions of ChatGPT and Claude for helping develop and refine the finite-$N$ Peierls argument. The authors are supported by the Simons Foundation International through the Simons Investigator program.

\bibliography{bibliography}

\onecolumngrid


\appendix

\section{$q$ at finite $N$}

\label{sec:qbound}

\begin{figure}[htbp]
\centering
\includegraphics[width=0.7\textwidth]{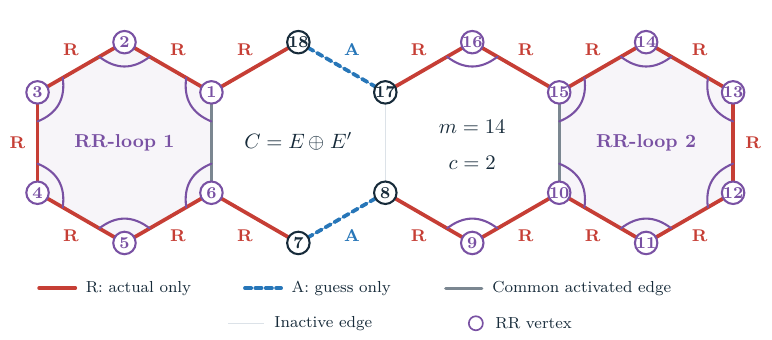}
\caption{One disagreement loop $C$ with two disjoint $RR$-loops. The red-and-blue boundary follows vertices $1,2,\ldots,18,1$. Red $R$ edges belong only to $E$, blue $A$ edges only to $E'$, and thick gray activated edges belong to both. The hexagons on vertices $1$--$6$ and $10$--$15$ close through the gray edges $(1,6)$ and $(10,15)$, which do not belong to $C$. Purple curves indicate the chosen singlet pairs following each hexagon and at the additional $RR$ vertices $9$ and $16$. The thin gray edge $(8,17)$ is inactive. This configuration has probability $N^{-24}$: an enhancement $N^4$ over $N^{-28}$.}
\label{fig:correlations}
\end{figure}

We now estimate and bound $q(N)$ in \Eq{eq:joint-hiding} for $\Sp(N)$ with $N\ge4$. Fix the actual activated edges $E$ and $m$ specified $RR$ vertices of a candidate disagreement loop $C$. We bound the probability that all these vertices hide a pair of fundamentals, conditioned on $E$ and leaving outcomes at all other vertices unrestricted. Our conservative estimate consists of a baseline pairing contribution and two corrections:

\begin{itemize}
\item \textbf{Baseline pairing contribution: $(3/N^2)^m$.}
Choose one pair of activated edges at each $RR$ vertex and project their two fundamentals into a singlet. There is one possible pair at a two-edge vertex; and three possible pairs at a three-edge vertex.
Thus there are {\it at most} $3^m$ choices overall. 
This gives an upper bound for the baseline contribution $\left(\frac{3}{N^2}\right)^m$.

\item \textbf{Maximum $RR$-loop enhancement: $(N^2)^{m/6}$.}
An \emph{$RR$-loop} is a closed loop of activated edges whose vertices are all selected $RR$ vertices of $C$. It may include activated edges common to $E$ and $E'$, so it need not follow $C$ throughout; see \figref{fig:correlations}.

An enhancement requires both an $RR$-loop and chosen pairs that follow it: at every vertex, the pair must consist of the two incident $RR$-loop edges. 
Around a closed $RR$-loop, after the preceding singlet projections succeed, the last pair is already a singlet. Its projection succeeds with certainty, giving an enhancement $N^2$. If there are in total $c$ $RR$-loops within a disagreement loop, the probability of these pair projections is
\beqn
N^{-2m}(N^2)^c=N^{-2m+2c}.
\label{eq:contraction-probability}
\eeqn

On the honeycomb torus with $L\ge6$, each closed component has at least six vertices; the shortest is a hexagon. 
\beqn
c\le\frac m6,
\qquad
(N^2)^c\le(N^2)^{m/6},
\qquad
N^{-2m+2c}\le N^{-5m/3}.
\label{eq:girth-bound}
\eeqn
For a conservative uniform bound, we grant \emph{every arrangement of the $m$ $RR$ vertices and every choice of pairs} the maximum enhancement $(N^2)^{m/6}$. 

\item \textbf{Projector-overlap factor: $(1-2/N)^{-m}$.}

At a three-edge vertex, the three pair-singlet subspaces are not orthogonal. We now derive the correction needed to bound the projector onto their combined fundamental sector by the sum of the three pair projectors.

For an arbitrary vector $v$ in the fundamental, the three cyclic pair-singlet embeddings have components
\beqn
\frac{1}{\sqrt N}J_{ab}v_c,
\qquad
\frac{1}{\sqrt N}J_{bc}v_a,
\qquad
\frac{1}{\sqrt N}J_{ca}v_b,
\label{eq:three-pairs}
\eeqn
where $J$ is the invariant symplectic form, normalized by $J^\dagger J=\one_N$. Each embedding preserves the norm. The overlap between two distinct cyclic embeddings is $-1/N$ times the inner product of their free vectors, giving the Gram matrix
\beqn
\begin{pmatrix}
1&-1/N&-1/N\\
-1/N&1&-1/N\\
-1/N&-1/N&1
\end{pmatrix}\otimes\one_N.
\label{eq:Gram}
\eeqn
Its eigenvalues are $1-2/N$ and $1+1/N$, the latter repeated twice in the three-dimensional pairing space. For $N>2$, all are positive, so the three pair-singlet subspaces together span a space of dimension $3N$.

These subspaces exhaust the fundamental sector of $F^{\otimes3}$. 
Let $P_{12},P_{23},P_{31}$ project onto the three pair-singlet subspaces, and let $Q$ project onto their combined fundamental sector. The nonzero eigenvalues of $P_{12}+P_{23}+P_{31}$ equal those of the Gram matrix in \Eq{eq:Gram}. Consequently,
\beqn
Q\le\frac{P_{12} + P_{23} + P_{31}}{1-2/N}.
\label{eq:projector-bound}
\eeqn
At a two-edge vertex, the hidden-pair projector is simply the unique pair-singlet projector, so no correction is needed. Applying the correction also at two-edge vertices gives the conservative uniform factor
\beqn
\frac1{(1-2/N)^m}.
\label{eq:overlap-factor}
\eeqn

\end{itemize}

Since the required local projectors act at distinct vertices, their bounds can be multiplied. Expanding the product of pair-projector sums yields at most $3^m$ terms, each a joint pair-projection probability bounded by \Eq{eq:girth-bound}. Combining the three factors gives
\beqn
\Pr(\text{all }m\text{ required vertices hide}\mid E)
&\le&
\underbrace{\left(\frac3{N^2}\right)^m}_{\text{baseline pairing}}
\underbrace{(N^2)^{m/6}}_{\text{maximum }RR\text{-loop enhancement}}
\underbrace{\frac1{(1-2/N)^m}}_{\text{projector bound}}
\nonumber\\
&=&
\left(\frac3{1-2/N}\,N^{-5/3}\right)^m.
\label{eq:three-factor-hiding}
\eeqn
Thus \Eq{eq:joint-hiding} holds uniformly in $E$ and in the choice of required vertices, with
\beqn
q(N)=\frac3{1-2/N}\,N^{-5/3}
=\frac{3N^{1/3}}{N(N-2)},
\qquad N\ge4.
\label{eq:qN}
\eeqn

The exceptional case $N=2$, for which the Gram matrix is singular and $F^{\otimes3}$ contains only two fundamental copies, requires a separate bound. It lies outside the sufficient range $N\ge40$ established in the main text. The same argument can be made for $\SU(N)$. In that case we find $q(N) = \frac{2 N^{1/3}}{N(N-1)}$ as long as $N\geq 6$. Eq.~\ref{eq:numerical-criterion} then gives the sufficient decodability bound $N\geq 31$ as stated in the main text.

\end{document}